\documentclass[9pt,twocolumn,twoside]{opticajnl}
\journal{opticajournal} 

\setboolean{shortarticle}{true}

\usepackage{lineno}
\usepackage{upgreek}
\usepackage{xcolor}

\title{Sensing the vibration of non-reflective surfaces with 10-dB-squeezed-light enhancement}

\author[1]{Pascal Gewecke}
\author[1]{Jascha Zander}
\author[1]{Roman Schnabel}

\affil[1]{Institut f{\"u}r Quantenphysik \& Zentrum f{\"u}r Optische Quantentechnologien, Universit{\"a}t Hamburg, Luruper Chaussee 149, 22761 Hamburg, Germany}

\affil[*]{roman.schnabel@uni-hamburg.de}

\begin{abstract}
Laser light with squeezed quantum uncertainty is a powerful tool for interferometric sensing. A routine application can be found in gravitational wave observatories. A significant quantum advantage is only achievable if a large fraction of the photons are actually measured.  For this reason, quantum-enhanced vibrational measurements of strongly absorbing or scattering surfaces have not been considered so far.
Here we demonstrate the strongly quantum-enhanced measurement of the frequency characteristics of surface vibrations in air by measuring the air pressure wave instead. Our squeezed laser beam, which simply passes the vibrating surface, delivers a sensitivity that an ultra-stable conventional light beam in the same configuration can only achieve with ten times the power. The pressure amplitude of a ultrasonic wave of just 
{0.12\,mPa$/\!\sqrt{\rm \mathbf{Hz}}$} was clearly visible with a spatial resolution in the millimetre range and a 1\,kHz resolution bandwidth. We envision applications in sensor technology where distant, highly absorbing or optically inaccessible surface vibrations in air are to be measured with limited, e.g.~eye-safe, light powers.
 \end{abstract}

\setboolean{displaycopyright}{false} 

\begin{document}
\sloppy
\maketitle

\section{Introduction}

Laser interferometric methods allow surface vibrations to be analyzed without mechanical contact. 
Heterodyne or homodyne laser Doppler vibrometry (LDV) usually analyze surface vibrations by measuring the phase modulations of back-reflected or back-scattered laser light. The power of the collected light sets an upper limit to the signal-to-noise ratio. For significant scattering, the shot noise is even overshadowed by speckle noise. Strong absorption by the vibrating surface can lead to additional systematic errors due to damage or thermal deformation. 

A significant enhancement of LDV by quantum correlated light can only successfully increase the signal to noise ratio if the measurement is quantum noise limited and if a significant fraction of the photons that carry the signal is actually measured \cite{Schnabel2017}. The most successful approach of sensing with quantum correlated light is the combination of a single quasi-monochromatic (`homodyne') carrier light with the signal sideband spectrum initially in a squeezed vacuum state \cite{Caves1981,Xiao1987,Grangier1987,Meylahn2022,Zander2023}.
If 50\% of the photons are photo-electrically converted squeezed light can still double the ratio of the signal power and the photon shot noise variance. If only 10\% of the photons are measured, the potential improvement drops to less than 5\% \cite{Schnabel2017}. Current gravitational wave (GW) observatories use squeezed vacuum states to achieve sensitivities that would not be possible without this technique \cite{LSC2011,Grote2013,Tse2019,Acernese2019,GWo3a2021,GWo3b2023}. Up to 75\% of the photons are converted into an identical number of photo electrons, resulting in a four times better ratio (6\,dB) of signal power to quantum noise variance \cite{Lough2021} using the automated `squeeze laser' with fast lock re-acquisition presented in \cite{Vahlbruch2010,Khalaidovski2012}. Squeezed states of light were also used to improve biological and micromechanical displacement measurements \cite{Taylor2013,Hoff2013}. Further references can be found in \cite{Lawrie2019}.

If the vibrating surface absorbs almost all photons or is shadowed, i.e.~not optically accessible, the described (surface back-scatter) 
LDV becomes inefficient or even impossible. In the first case, the laser power could be increased, but this would also increase the heat input into the oscillator. In extreme cases, this would change the oscillation properties and cause systematic measurement errors. The second case involves a sound source that is invisible and emits sound via a labyrinth of air ducts.  A concrete example is the human vocal chords, which are not optically accessible, but the airborne sound produced is.

If the surface vibrates in ambient air, another gas in the viscous flow regime or a liquid, acousto-optic laser measurements are powerful alternatives. Here, the laser beam simply passes the vibrating surface. Two different approaches exist. 
The older one uses the deflection of a laser beam at the phase grating created by the acoustic waves \cite{Phariseau1956,Klein1967,Schroedel2023}. 
Due to `acousto-optic modulation', the light's frequency is shifted by the frequency of the acoustic wave.
In this case, the optical beam diameter has to be larger than the acoustic wavelength.\\ 
The newer one is known as `light refractive tomography'. It uses an optical beam diameter that is smaller than the acoustic wavelength, which allows the tomographic reconstruction of the sound wave \cite{Jia1993,Matar2000,Zipser2002,Harvey2006,Rupitsch2014}.
The laser beam picks up a phase modulation due to the oscillating refractive index caused by the sound wave, which is then detected interferometrically. A strong signal is only created at locations along the beam path where the beam diameter reaches a value smaller than the acoustic wavelength. For instance, a 10\,MHz ultrasonic wave has a wavelength of 150\,$\upmu$m in water, which can easily be resolved in three dimensions \cite{Almqvist1999}. 
The measurement concept of light refractive tomography is useful even without the tomography aspect. It can be used to measure the frequencies and mechanical quality factors of vibrating surfaces if they are strongly absorbing, i.e.~black, and if they are not optically but acoustically accessible due to other mechanical components. We speak here of
`acousto-optic vibration sensing (VS)'. 
The tomographic aspect of this measurement concept comes into play again, when a large number of different surface vibrations are superimposed forming one complex acoustic signal. The tomographic measurement enables the decomposition into components, including the localisation of the vibrating components. 

Acousto-optic {VS} can be enhanced by squeezed light, similar to GW detection, which has not been demonstrated so far. The preconditions are (i) photo shot noise limited sensitivity and (ii) the photo-electric detection of a significant fraction of those photons that actually propagated through the acoustic wave. A strong motivation for squeezed light enhancement is given if the light power cannot be further increased easily to benefit from an improved signal to shot-noise ratio, see Eq.\,(\ref{eq:3}). This is the case if the laser light is already acting back on the acoustic wave, 
{e.g.~because the optical absorption heats the air,} 
or simply if eye-safe light power is mandatory \cite{Zander2023}.

Here we report on the elevation of acousto-optic {VS} into the field of quantum sensing with quantum correlated light. We measure the vibration frequency of a surface without shining light onto it, concretely, we measure the frequency of an ultrasonic transducer in ambient air by laser-interferometric homodyne detection of the sound wave with a nonclassical sensitivity improvement of 10\,dB. The same sensitivity improvement can semi-classically only be achieved by using a tenfold light power. Our approach is independent from the roughness and absorption of the transducer surface and allows the full advantage of squeezed light. 

\section{Setup}
\begin{figure}[ht!!!]
        \center{\includegraphics[width=80mm]{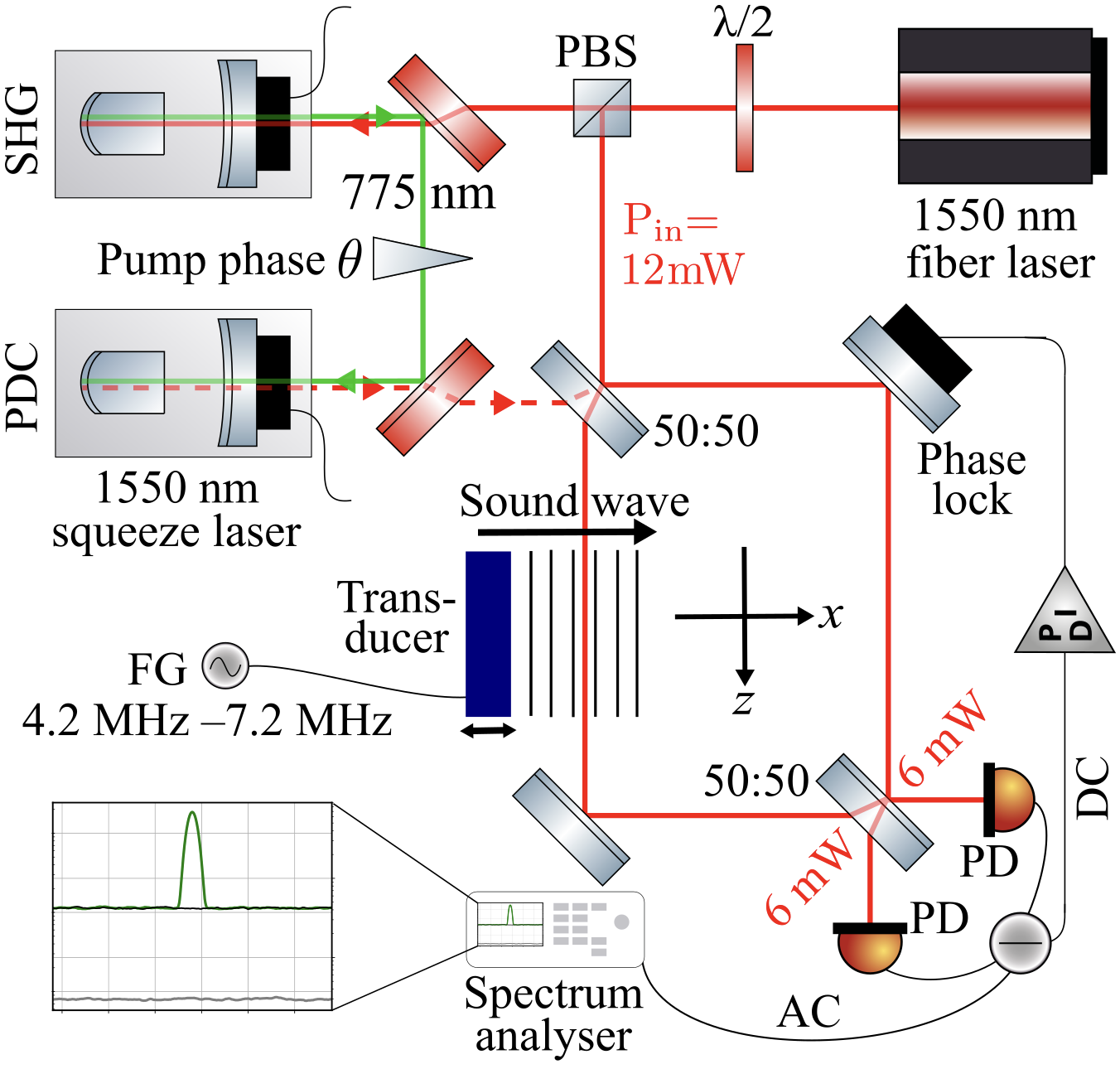}}
        \caption{\textbf{Schematic of the experiment} -- The sound wave from an ultrasonic transducer vibrating at a variable frequency at a few MHz was detected with 12\,mW quasi-monochromatic laser light at 1550\,nm combined with a second beam with the corresponding MHz-sideband spectrum in a squeezed vacuum state. The latter was produced by successive second harmonic generation (SHG) and cavity-enhanced parametric down-conversion (PDC). 
        The setup corresponded to a quantum-enhanced Mach-Zehnder interferometer electro-optically controlled to the usual mid-fringe operation. The DC output of the balanced detection was used for the electro-optical control. The AC output contained the information about the frequency of the transducer. PBS: polarising beam splitter; {$\lambda\!/\!\!\;2$}: half-wave plate; FG: frequency generator; PD: photo diode.  
        \label{fig:1}}
\end{figure}

Figure\,\ref{fig:1} shows a sketch of our squeezed-light-enhanced acousto-optic {VS} setup. About $700\,$mW of quasi-monochromatic light at 1550$\,$nm was generated by a fiber laser provided by \emph{NKT\,Photonics}. $12\,$mW was tapped at a polarizing beam splitter acting as the carrier light, while the other part was used to produce a beam with its MHz sideband spectrum in a squeezed vacuum state \cite{Wu1986,Vahlbruch2016,Schnabel2017}. Two identical 9.3$\,$mm-long periodically poled, quasi-phase-matched potassium titanyl phosphate (PPKTP) crystals inside optical resonators first produced about 0.3\,W of 775\,nm via second harmonic generation (SHG), which was then used to pump parametric down-conversion (PDC) slightly below its oscillation threshold for optimal squeezed light generation \cite{Wu1986,Vahlbruch2016,Schnabel2017}.

The squeezed beam and the 12\,mW-beam were overlapped on a first balanced beam splitter with a differential phase $\theta$ such that the interferometer's differential phase signal showed a squeezed quantum uncertainty. Only one of the beam splitter output beams passed the ultrasonic wave of a piezo-electric transducer and picked up a monochromatic MHz phase modulation. The two beams were recombined at the second balanced beam splitter forming a Mach-Zehnder interferometer. The interference contrasts at both beam splitters were above 99\%. The interferometer was operated at the mid-fringe condition, i.e.~the final beams had the same power of 6$\,$mW and were measured with PIN photo diodes (PDs) with a quantum efficiency of approximately 99\%.
The differential photo current was fed into a transimpedance amplifier. The AC part of the produced signal was read out on a spectrum analyzer, while the DC part was used for the length stabilization of the interferometer. The control loop actuator was a piezo-positioned steering mirror inside the interferometer, see Fig.\,\ref{fig:1}.

The piezo-electric transducer used for all measurements was a dual element transducer from \emph{Smart Sensor}. It had a diameter of $15\,$mm and could produce continuous ultrasonic sound waves between $4.2\,$MHz and $7.2\,$MHz. Its resonance frequency was $5.2\,$MHz, which resulted in an acoustic wavelength of 61\,$\upmu$m. The waist of the laser beam $w_0$ in front of the transducer was focused down to $w_0 = 31\,\upmu$m. Its position could be moved in all directions enabling tomographic measurements on the sound wave amplitude.

\section{Measurement Analysis}

\begin{figure}[ht!!!]
        \center{\includegraphics[width=87mm]{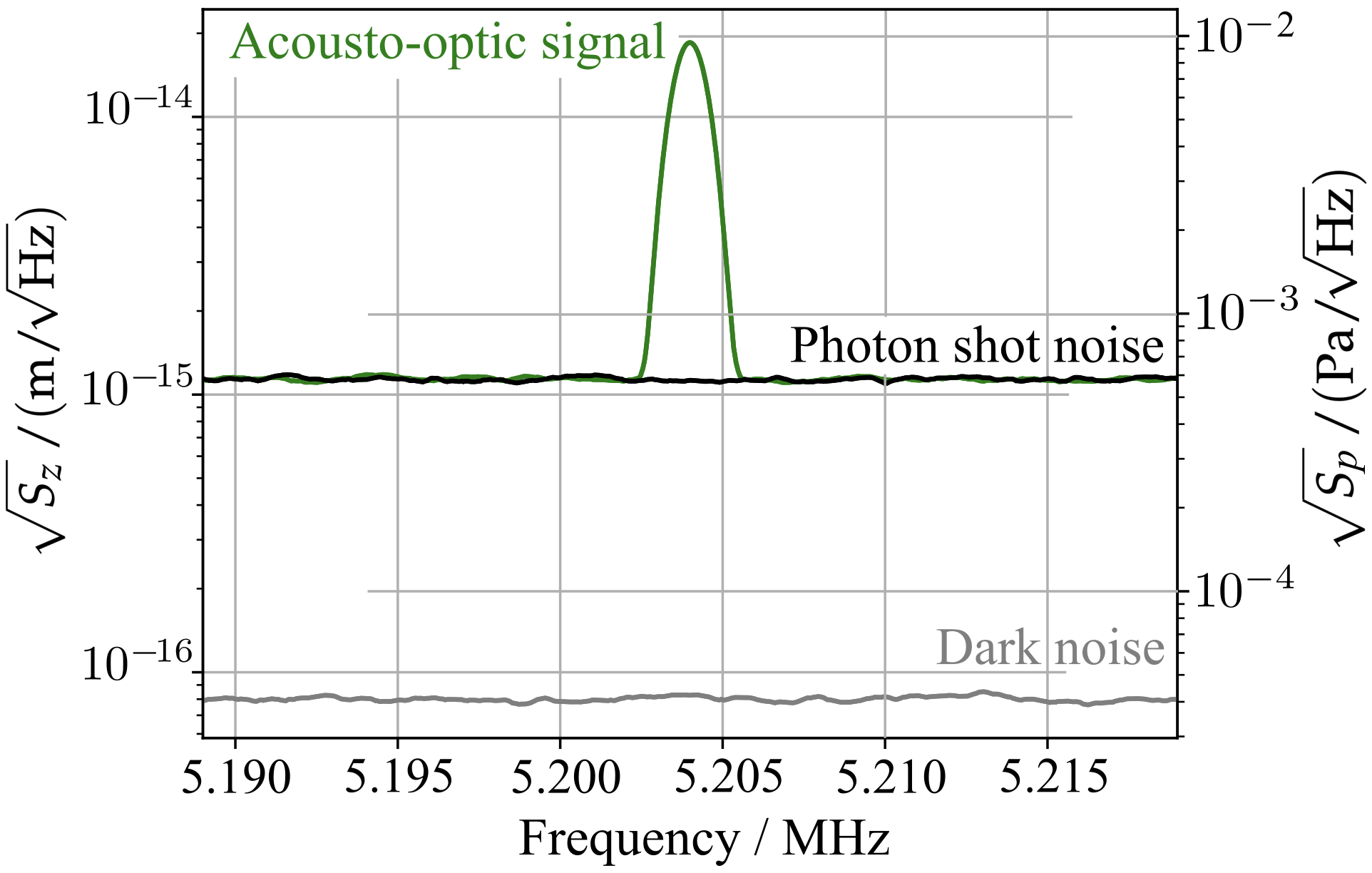}}
        \caption{\textbf{Example of a strong acousto-optic {VS} signal} -- Here the transducer was driven with a relatively high voltage. Shown is the amplitude spectral density normalized to effective optical path length change according to Eq.\,(\ref{eq:3}). The total power in the interferometer was $P_{\textrm{in}} = 12\,$mW. The photon shot noise was measured while the laser beam was shielded against the operating transducer. The spectra were measured with a resolution bandwidth (RBW) of 1$\,$kHz, a video bandwidth (VBW) of 10$\,$Hz and averaged 30 times. The width of the acousto-optic signal was limited by the RBW. The normalization of the y-axis to ${\rm Pa}/\!\!\sqrt{\rm Hz}$ was carried out using Eq.\,(\ref{eq:4}).
        }
\label{fig:2} 
\end{figure}

Figure\,\ref{fig:2} shows a strong acousto-optic signal measured with the setup in Fig.\,\ref{fig:1}. 
A sound wave at frequency $f$ corresponds to an harmonic oscillation of the air pressure, which leads to an oscillation of the refractive index 
\begin{align} \nonumber
n(x,y,z,t) 	&= n_0 +  \Delta n(x,y,z) \cdot {\rm cos} [2\pi ft+\phi(x,y,z)] \\
		&= n_0 + \frac{\delta n}{\delta p} \cdot \Delta p(x,y,z) \cdot {\rm cos}[2\pi ft+\phi(x,y,z)] \, ,
\label{eq:1}
\end{align}
where {$p$ is the air pressure,} $n_0$ the refractive index of the air in the absence of sound waves, and $\delta n \!\!\;/\!\!\; \delta p$ is the piezooptic coefficient of the air.
At a temperature of 20$^\circ$C, an ambient pressure of 1013.25\,mbar, a relative humidity of 40\%, and a carbon dioxide content of 0.045\% its value is $\delta n \!\!\;/\!\!\; \delta p =2.072 \cdot 10^{-9} \,\textrm{Pa}^{\!-1}$ \cite{Rupitsch2014}.
The amplitude of the optical path length oscillation of a strongly focussed beam of light propagating parallel to the sound wave fronts is given by
\begin{align} 
\label{eq:2}
\Delta L(x,y,z) \approx z_{\rm M} \cdot \Delta n(x,y,z)\, ,
\end{align}
where $z_{\rm M}$ is the effective length over which the radius of the laser beam is small enough to capture the {{\it full} modulation depth} 
of the refractive index.  
Our laser beam had an almost perfect transverse TEM{\small 00} Gaussian mode, and approximately 2\,mm before and after the waist position the beam radius reached approximately 0.7 of the acoustic wavelength, beyond which the refractive index modulation is almost fully washed out.
{The beam's focus had a one-sigma intensity diameter of approximately 44\,$\upmu$m, which corresponded to 72\% of the acoustic wavelength. This means that the depth of the refractive index modulation is also strongly washed out even {\it within} the 4 mm propagation distance.
As a very rough estimate, we use an effective value $z_{\rm M} \approx 1$\,mm, within which the modulation is  not washed out at all.}

The interferometer of our setup transforms the differential arm length modulation into a modulation of the differential photo-electric current whose normalized amplitude spectral density is shown in Fig.\,\ref{fig:2}. Also shown is the photon shot noise level of the detected 12\,mW of laser light. It was measured with the same setup while the laser beam was shielded against the operating transducer. The absence of any signal ensured that the signal was not caused by side effects such as electromagnetic stray fields. The dark noise was measured while all light beams were blocked. 
We normalized the peak of the acousto-optic signal to the unit ${\rm Pa}/\!\sqrt{{\rm Hz}}\,$ by the well-known normalization of interferometric photon shot noise to optical path length change $z$ in the unit ${\rm m}/\!\sqrt{{\rm Hz}}\,$ \cite{Saulson1994,Schnabel2017,Genovese2021} following \cite{Zander2023}
\begin{equation}
\label{eq:3}
\sqrt{S_z} = \sqrt{\frac{h c \lambda}{2 \pi^2  P_{\textrm{in}}}}  \,,
\end{equation}
where $h$ is Planck's constant, $c$ is the speed of light, $\lambda$ is the optical wavelength of quasi-monochromatic carrier light and $P_{\textrm{in}}$ the total light power in both interferometer arms. The equation neither includes dark noise nor the influence of potential imperfections. We estimated the optical power loss of our experiment to about 2\% taking into account propagation losses as well as the imperfect quantum efficiency of our photon diodes of ($99 \pm 1$)\%. After adding the dark noise level we found our photon shot noise level slightly above one ${\rm fm}/\!\sqrt{{\rm Hz}}$.
The acousto-optic peak thus corresponds to a measured optical path length change of approximately $2 \cdot 10^{-14} {\rm m}/\!\sqrt{{\rm Hz}} \cdot \sqrt{\rm kHz} \approx 0.6\,$pm, where the $\rm kHz$ corresponds to the measurement's resolution bandwidth.

The amplitude spectral densities in Fig.\,\ref{fig:2} can be normalized to ${\rm Pa}/\!\sqrt{{\rm Hz}}\,$ by
\begin{equation} 
\label{eq:4}
 \sqrt{S_p} = \frac{\sqrt{S_z}}{z_{\rm M}} \Bigl(\frac{\partial n}{\partial p}\Bigr)^{\!\!-1}\,, 
 \end{equation}
where $z_{\rm M} \approx 1\,$mm is the effective measurement length and $\delta n \!\!\;/\!\!\; \delta p = 2.072 \cdot 10^{-9} \,\textrm{Pa}^{\!-1}$ \cite{Rupitsch2014}. With these calculations, the height of the peak in Fig.\,\ref{fig:2} corresponds to approximately $9\,\textrm{mPa} / \!\sqrt{\textrm{Hz}}$ minus a pressure noise equivalent of $0.55\,\textrm{mPa} / \!\sqrt{\textrm{Hz}}$ from photon shot noise.\\

\begin{figure}[ht!!!]
        \center{\includegraphics[width=83mm]{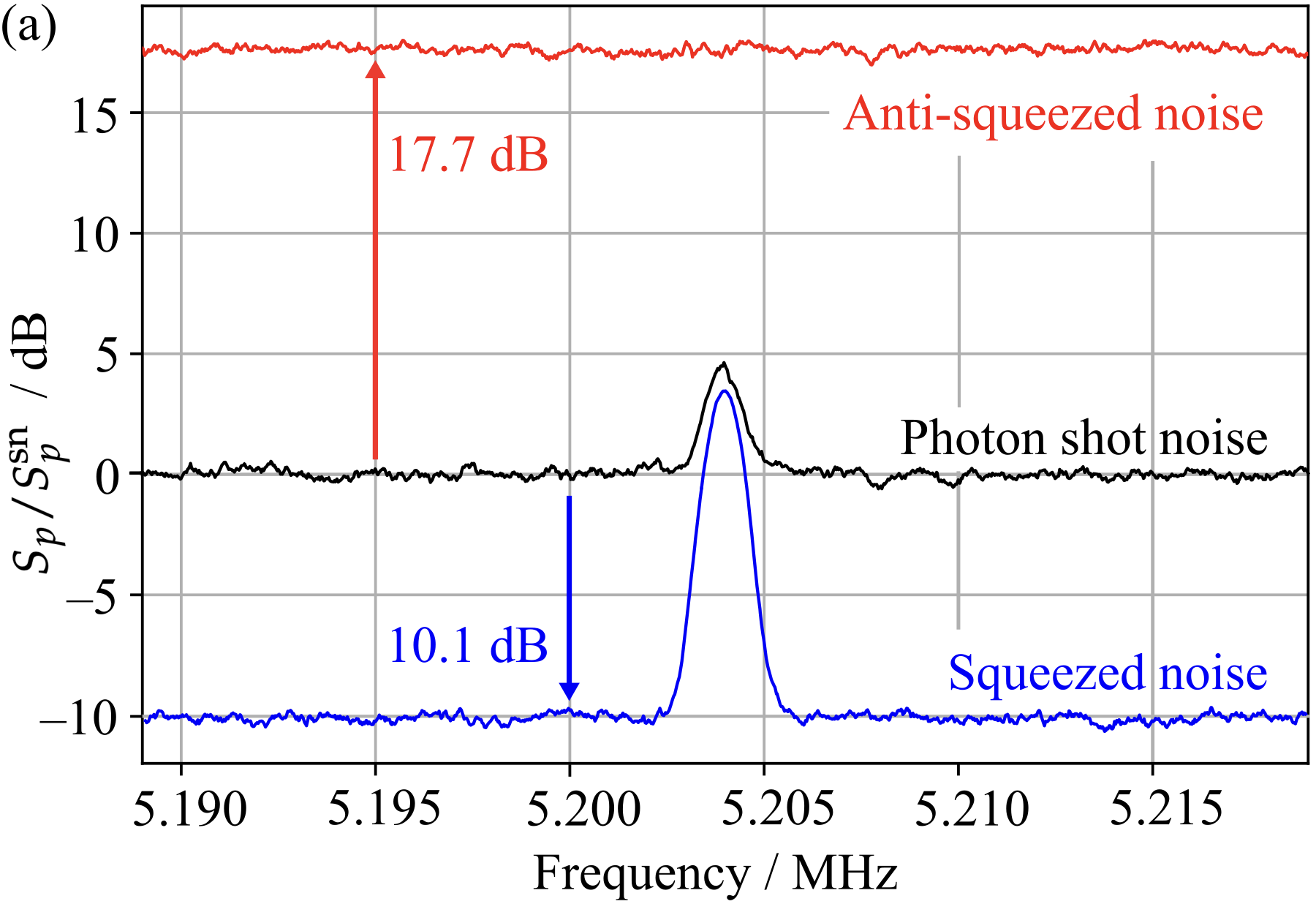}}
        \center{\includegraphics[width=83mm]{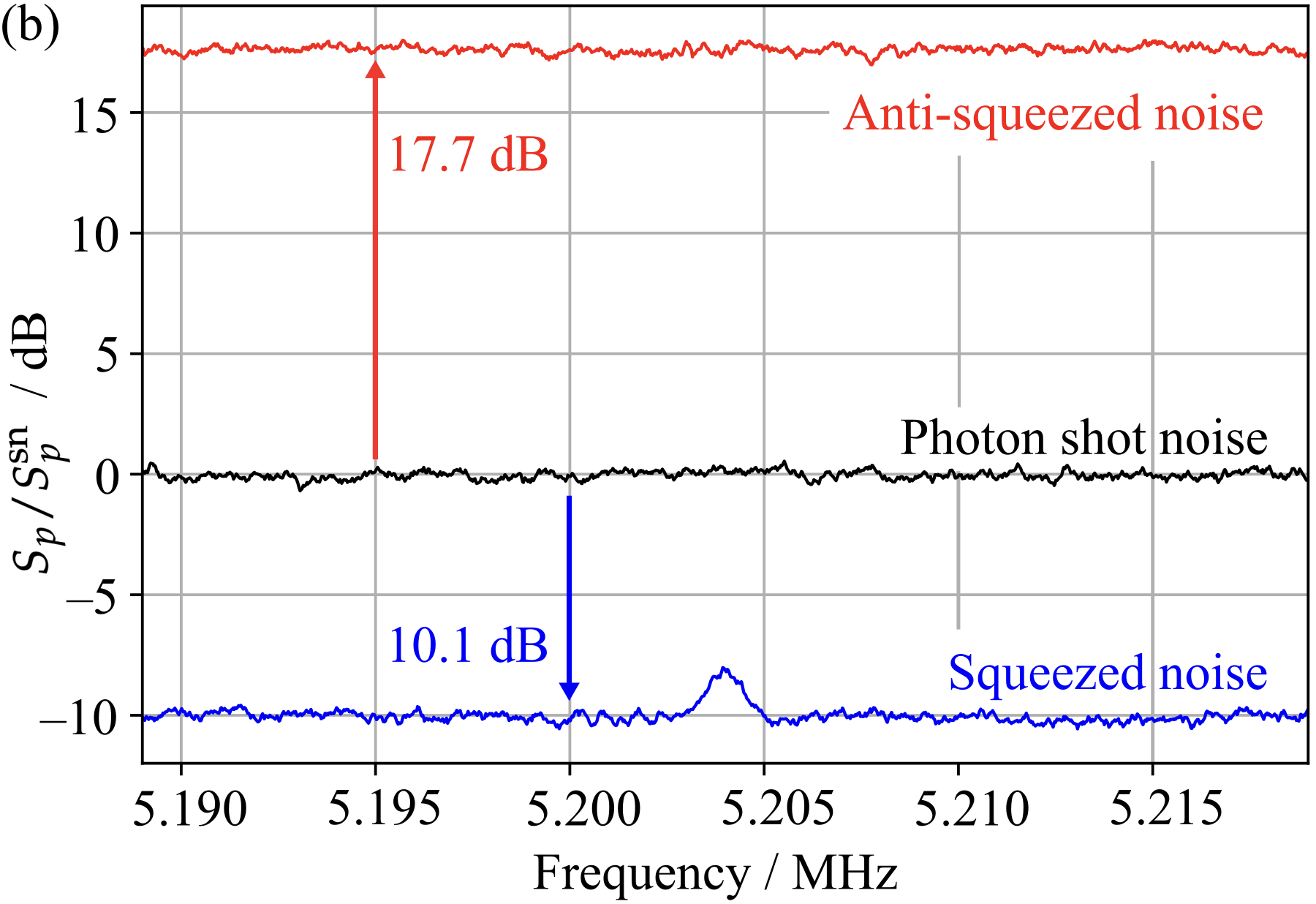}} 
        \caption{\textbf{Squeezed-light enhanced acousto-optic {VS}} -- Shot-noise normalized power spectral densities characterizing two ultrasound measurements. The shot noise levels correspond to that of Fig.\,\ref{fig:2} ($P_{\textrm{in}}\!=\!12\,$mW, RBW: 1$\,$kHz, VBW: 10$\,$Hz, 30 times averaging). (a) Converting the peak height to the linear scale and subtracting the quantum noise result in a peak 1.5 times the photon shot noise, i.e.~approximately $0.8\,\textrm{mPa} / \!\sqrt{\textrm{Hz}}$. The even lower ultrasound signal in (b) was only visible in the squeezed spectrum and corresponded to approximately $0.12\,\textrm{mPa} / \!\sqrt{\textrm{Hz}}$. The anti-squeezed noise completely covered both signal strengths and was only measured once. It was produced with a change of the pump phase by $\Delta \theta = \pi$. (The dark noise level was not subtracted.)}
        \label{fig:3}
\end{figure}
\begin{figure}[ht!!!]
        \center{\includegraphics[width=85mm]{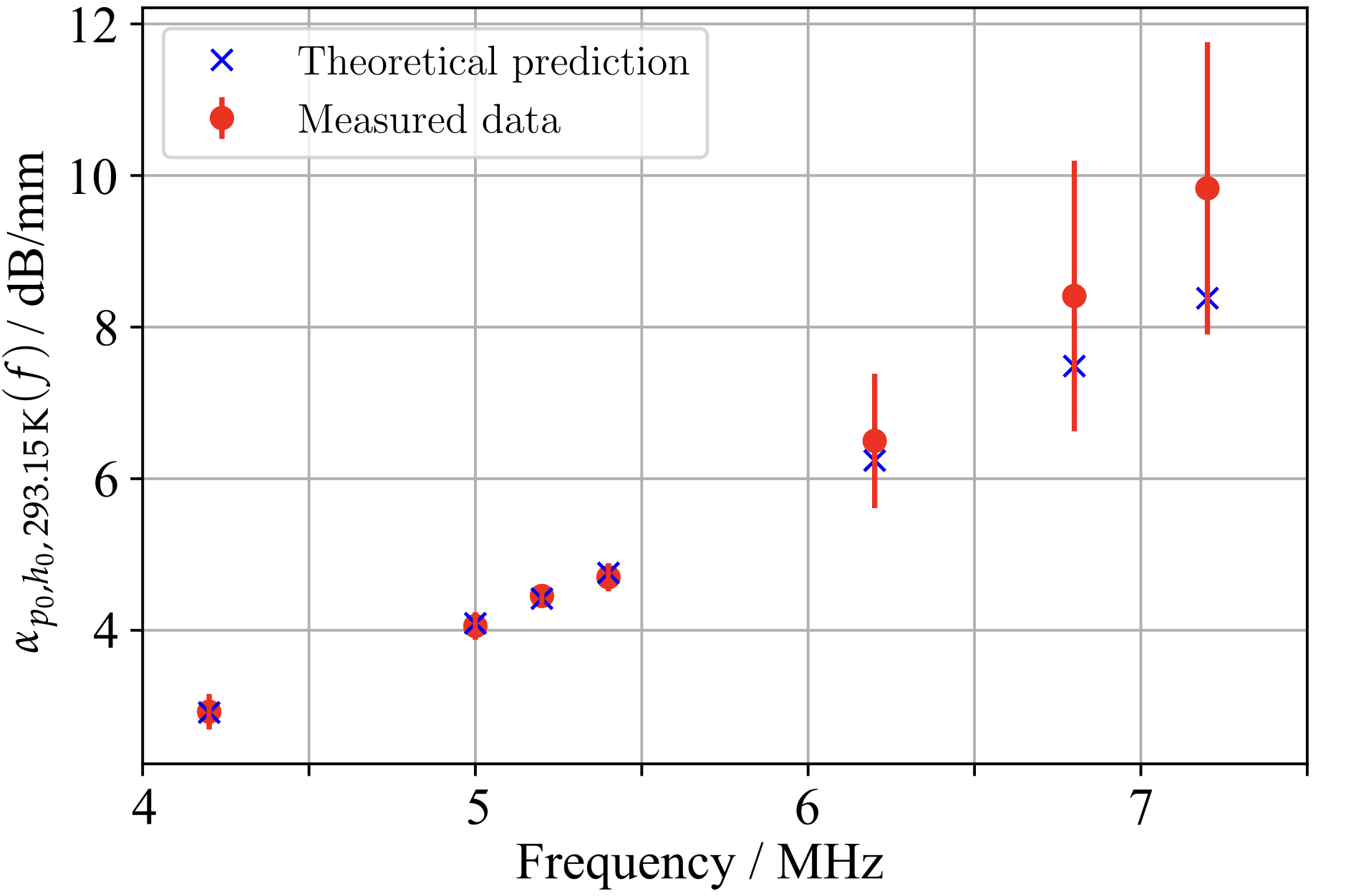}}
        \caption{\textbf{Absorption coefficient of air versus frequency} -- \\
        Each of the measurement points (circles) was derived from a series of 
        {(shot-noise-limited)} 
        measurements with different distances between the transducer and the laser beam. The error bars for the 
        first four frequencies are very small, as ten distances between zero and seven millimetres could be realised. 
        At the highest frequency, however, the signal was already too weak at distances greater than 0.7\,mm, which 
        was mainly due to our source. Overall, the measurement results agree well with Eq.\,(\ref{eq:5}) (crosses). 
        The temperature and pressure of the air were around 20$^\circ$C and 1\,bar.}
        \label{fig:4} 
\end{figure}
Fig.\,\ref{fig:3} presents our measurements demonstrating squeezed-light enhanced acousto-optic laser Doppler vibrometry of a body vibrating in ambient air without shining light onto it. Here, we used a pressure amplitude of the acoustic wave that was about a factor of 50 lower than that in Fig.\,\ref{fig:2}.
Consequently, the photon shot noise level completely buried the acousto-optic signal. The light power was again 12\,mW. By injecting a `beam' with a MHz sideband spectrum in a larger than 10-dB squeezed vacuum state into the open interferometer port, the small acousto-optic {VS} signal became visible.
Interference contrasts of greater than 99\% at both beam splitters and the photo diode quantum efficiencies of approximately 99\% as well made it possible to maintain a squeeze factor of 10\,dB in the measured power spectral densities. Please note that the traces in Fig.\,\ref{fig:3} are normalized to the variance of the photon shot noise. They are thus proportional to the units m$^2\!/$Hz and Pa$^2\!/$Hz. 
The upper trace was taken with anti-squeezed quantum noise 
{produced with a change of the pump phase by $\Delta \theta = \pi$.}
We used the value of the anti-squeezed noise power to estimate the total optical loss on the squeezed vacuum state \cite{Vahlbruch2016}. We derived a total photon loss value of 7.8$\%$, which also includes inefficiencies in squeezed light generation and measurement. In principle, this value allows for the measurement of $10\cdot {\rm log}\,0.078 \approx 11$\,dB of squeezing at parametric oscillation threshold \cite{Schnabel2017}. However, we used a slightly lower gain to keep the cavity length control stable and phase noise at a negligible level \cite{Franzen2006}.

\begin{figure}
        \center{\includegraphics[width=85mm]{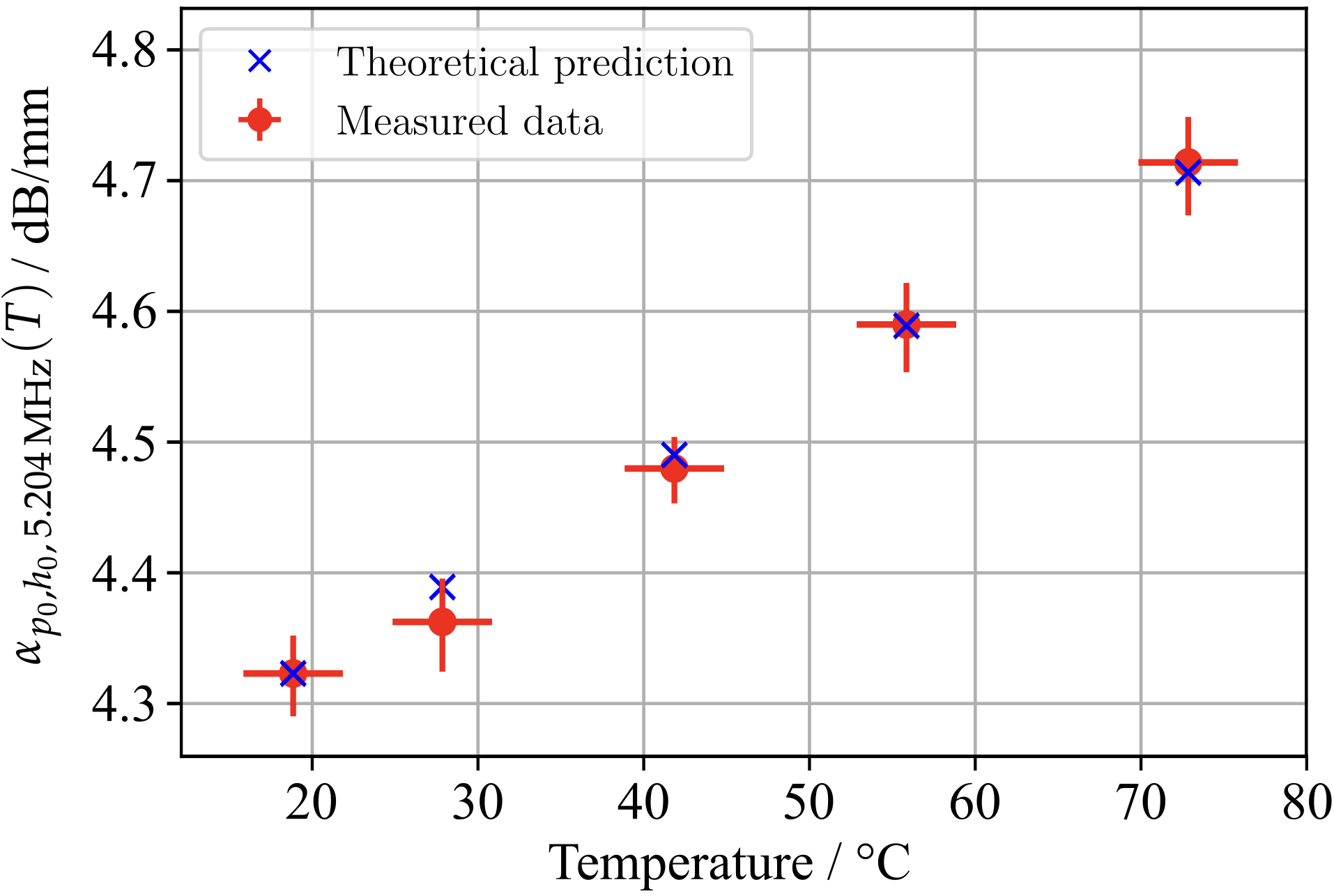}}
        \caption{\textbf{Absorption coefficient of air versus temperature} -- \\
        The absorption coefficient at different temperatures at the fixed frequency $f = 5.204\,$MHz, measured close 
        to the source (circles) 
        {with shot-noise-limited light}. 
        The peak height was measured for each data point and then converted into the absorption coefficient. The 
        measurement results agree well with Eq.\,(\ref{eq:5}) (crosses). The pressure of the air was around 1\,bar.}
        \label{fig:5} 
\end{figure}

Ambient air has a rather high absorption of ultrasonic waves, which depends on frequency as well as temperature. The dependencies were studied extensively in \cite{Bond1992} by using two transducers, one emitting the waves and one receiving it. 
We used our setup to cross-check some of their findings and measured the change of the acousto-optic signal versus ultrasonic frequency in the range 4.2 to 7.2 MHz (Fig.\,\ref{fig:4}) and for the frequency of within the temperature range of 10 to 72$^\circ$C (Fig.\,\ref{fig:5}). We find very good agreement with theoretical models from literature.
The absorption coefficient of sound waves at MHz frequencies in ambient air with pressure $p_0=1013.25\,$mbar and humidity $h_0 = 40\%$ in dB$/$mm is given by \cite{ISO9613}
\begin{equation} 
\label{eq:5}
\alpha_{p_0,h_0}(T,f) \approx 15.895 \cdot 10^{-14} \Bigl( \frac{T}{293.15\,{\rm K}} \Bigr) f^2,\
\end{equation}
where $f$ is the ultrasonic frequency and $T$ the temperature of the ambient air.

Figure \ref{fig:4} shows the absorption coefficient of air for frequencies in the range between 4.2$\,$MHz and 7.2$\,$MHz, limited by the bandwidth of the transducer. For each frequency, we measured the peak height for different distances between the transducer and the laser beam. The data was converted into units of Watt, and an exponential fit was performed to the data, from which we extracted the absorption coefficient. The data is in very good agreement with the theoretical prediction done in Eq.\,(\ref{eq:5}).
Figure \ref{fig:5} shows the absorption coefficient of air versus temperature from 18 to 72$^\circ$C. To perform this measurement, we built an oven made out of a copper tube and heating wire coiled around it. The oven temperature was measured with a negative temperature coefficient thermistor (NTC) that was placed inside the oven. Again, our data matches well the theoretical description. 
Both series of measurements support the fact that the measured signals were generated by the acousto-optical interaction and at the same time demonstrate the high sensitivity of the setup.

\section{Conclusions}

{
Frequencies of surface vibrations are usually measured using the phase change of laser light that is reflected or scattered back from the vibrating surface. Commercial devices are called `Laser-Doppler-Vibrometers'. Highly absorbent and in particular geometrically concealed surfaces, however, set a limit to this approach. Improvement through quantum-correlated laser light cannot generally be widely used unless a significant proportion of the light is also measured. In this study we use laser light in such a way that it only penetrates the sound wave emanating from the surface. In this case, vibrations can also be measured from surfaces that completely absorb light or are not optically accessible. Our method allows the measurement of almost 100\% of the light, which enabled us to demonstrate a sensitivity improvement of 10\,dB using quantum correlated light.  A general limitation of our method is that the vibrating surface must be surrounded by a sufficiently high gas pressure. Based on typical ambient conditions, the signal strength is inversely proportional to the gas pressure \cite{Rupitsch2014}.}

In our experiment, we use laser light at 1550\,nm to measure the vibration of an ultrasonic transducer via the {5\,MHz pressure vibration in ambient air. The absorption of 1550 nm light by the air is negligible (even over a distance of several dozen metres), and we effectively use strongly squeezed states of light to improve measurement sensitivity to 10 dB below photon shot noise.
Only recently, such a high value was achieved for the first time (in homodyne interferometers) \cite{Zander2023,Heinze2022}. Squeezed states were applied to heterodyne interferometers but yielded much less squeezing in the measured data \cite{Li2015,Yu2023}, because overlapping the squeezed field with the measurement beam as well as additional optical components such as Bragg cells (acousto-optic modulators), wave plates and polarizing optics introduced additional optical loss due to absorption, scattering and reduced mode matching in the photoelectric detection schemes.
} 

Using {12}\,mW quasi-monochromatic carrier light with a MHz sideband spectrum in a squeezed vacuum state, we achieve a spectral amplitude noise density of 
{$0.12\,{\rm mPa}/\!\sqrt{\rm Hz}$ with a resolution bandwidth of 1\,kHz} 
and a spatial longitudinal resolution in the mm range. 
{The high RBW means that temporal changes in the signal on a millisecond scale could still be registered.}
{Our sensitivity could be further improved by increasing the light power and/or the squeeze factor, but also by focussing the laser light more strongly or by switching to a lower ultrasonic frequency, which increases the effective interaction length. Although our work was not aimed at maximising absolute sensitivity but at maximising enhancement by quantum correlations, our absolute result is slightly better than that of previous work carried out at advantageously lower frequencies. Refs. \cite{Jia1993,Matar2000} reported $\approx 0.3\,{\rm mPa}/\!\sqrt{\rm Hz}$ for ultrasound at 2\,MHz and $\approx 0.2\,{\rm mPa}/\!\sqrt{\rm Hz}$ below 1\,MHz, respectively.}

Our approach can also be applied to pressure waves at audible frequencies. Pressure waves in this range are much less absorbed by air, which means that the distance of the laser beam from the emitting surfaces can be many tens of metres. In addition, audible wavelengths are so long that the divergence of the laser beams becomes almost irrelevant and the interaction distances become much greater, which significantly increases signal strengths. If laser beams propagate over longer distances through air, laser safety is usually an issue. With a laser wavelength of 1550\,nm, eye safety is only guaranteed up to 10\,mW. Combining 10\,mW of carrier light with a 10\,dB squeezed sideband spectrum results in a sensitivity identical to that of 100 milliwatts of shot-noise-limited light. The quantum technology of squeezed light thus has a convincing motivation in acousto-optical {vibration sensing}.

\section{Backmatter}

\begin{backmatter}

\bmsection{Funding} 
This work was supported by the DFG under Germany's Excellence Strategy EXC 2121 Quantum Universe -- 390833306. 

\bmsection{Acknowledgments} 
We thank Mikhail Korobko, Christian Rembe, Jan S\"udbeck, and Mengwei Yu  for fruitful discussion.\\

\bmsection{Disclosures} 
The authors declare no conflicts of interest.

\smallskip

\bmsection{Data Availability Statement} 
The data that support the plots within this paper and other findings of this study are available from the corresponding author upon reasonable request.

\end{backmatter}

\bibliographystyle{plain}

\end{document}